\documentclass{Interspeech}
\usepackage{multirow}
\usepackage{mathrsfs}
\usepackage{hyperref}
\usepackage[subtle]{savetrees}

\newcommand{\specialcell}[2]{%
  \begin{tabular}[c]{@{}m{#1}@{}} \centering #2 \end{tabular}}

\interspeechcameraready

\title{StutterCut: Uncertainty-Guided Normalised Cut for Dysfluency Segmentation}

\author[affiliation={1}]{Suhita}{Ghosh}
\author[affiliation={2}]{Melanie}{Jouaiti}
\author[affiliation={1}]{Jan-Ole}{Perschewski}
\author[affiliation={1}]{Sebastian}{Stober}

\affiliation{Artificial Intelligence Lab}{Otto-von-Guericke-University}{Germany}
\affiliation{School of Computer Science}{University of Birmingham}{England}
\email{\{suhita.ghosh,jan-ole.perschewski,stober\}@ovgu.de, m.jouaiti@bham.ac.uk}

\usepackage{comment}

\begin{document}

\maketitle

\begin{abstract}
Detecting and segmenting dysfluencies is crucial for effective speech therapy and real-time feedback.
However, most methods only classify dysfluencies at the utterance level.
We introduce StutterCut, a semi-supervised framework that formulates dysfluency segmentation as a graph partitioning problem, where speech embeddings from overlapping windows are represented as graph nodes.
We refine the connections between nodes using a pseudo-oracle classifier trained on weak (utterance-level) labels, with its influence controlled by an uncertainty measure from Monte Carlo dropout.
Additionally, we extend the weakly labelled FluencyBank dataset by incorporating frame-level dysfluency boundaries for four dysfluency types.
This provides a more realistic benchmark compared to synthetic datasets.
Experiments on real and synthetic datasets show that StutterCut outperforms existing methods, achieving higher F1 scores and more precise stuttering onset detection.
\end{abstract}

\section{Introduction}

Speech dysfluencies disrupt the natural flow of speech and diminish communicative ability and quality of life for those who stutter~\cite{lian2024towards, craig2009impact}.
Although deep learning has shown promise in detecting dysfluencies at the utterance level~\cite{bayerl2023classification, changawala2024whister}, pinpointing the exact onset and boundaries remains understudied.
Such precise localisation is crucial for targeted therapy~\cite{freeburn2022speech}, real-time feedback in clinical assessment~\cite{laiho2022stuttering} or self-guided contexts~\cite{saldanha2023real,Belke2024}, and generating realistic synthetic data~\cite{zhou2024yolo}.
\begin{figure}[!t] 
\centering
  \includegraphics[width=\linewidth]{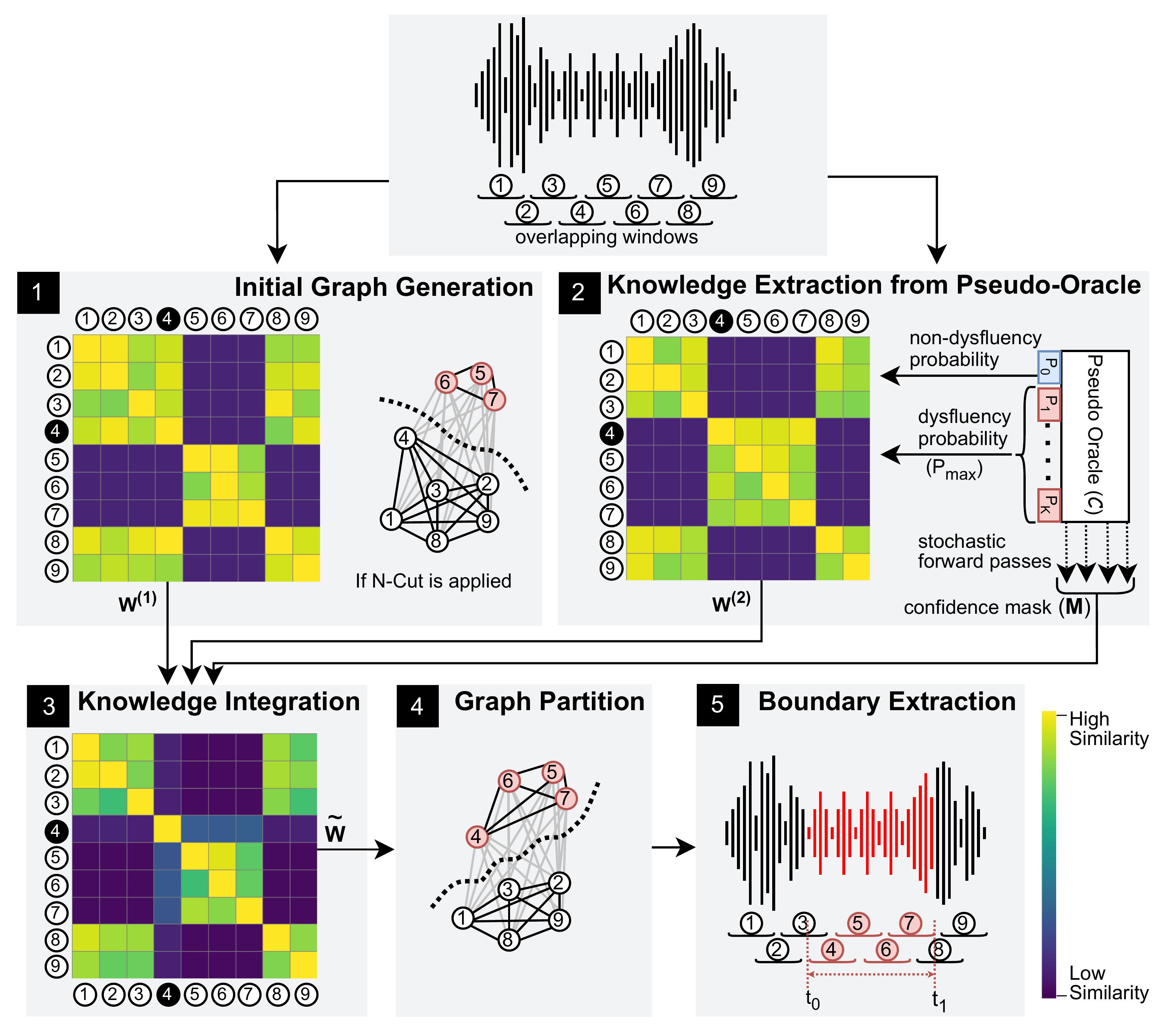}
  \vspace{-1.5em} 
  \caption{The input utterance is divided into overlapping windows, with embeddings from each window forming nodes in a graph. The connection strength of nodes is represented by colour intensity, with darker lines indicating stronger links. Dysfluent nodes and frames are highlighted in red. Notably, node 4, positioned at the boundary between fluent and dysfluent regions in the latent space, is initially misclassified as fluent but is correctly classified after integrating uncertainty-weighted classifier knowledge.}
  \label{fig:sc}
   \vspace{-2em}
\end{figure}

A key obstacle is the scarcity of real-world datasets with strong (frame-level) dysfluency labels.
Existing strong-labelled synthetic corpora, such as LibriStutter~\cite{libristutter} and VCTK-based dysfluent corpora~\cite{zhou2024yolo}, often lack the natural variability of real speech and introduce unnatural artefacts.
These include synthetic noise in prolongations~\cite{changawala2024whister}, robotic distortions~\cite{zhou2024yolo}, and overly rigid pauses in blocks.
In contrast, real stuttering exhibits subtle variations, such as fluctuating energy in repetitions and gradual transitions, which synthetic datasets fail to capture accurately.

Few methods address dysfluency segmentation.
The most recent, YOLO-Stutter~\cite{zhou2024yolo}, applies the well-known region-based segmentation framework YOLO~\cite{redmon2016you} to derive dysfluency boundaries.
It leverages a pre-trained VITS~\cite{kim2021conditional} model to align speech and text embeddings, followed by spatial and temporal aggregation.
However, its reliance on reference texts, which are often unavailable or inaccurate for atypical speech when using general ASR models such as Whisper~\cite{radford2023robust}, limits its real-world applicability.
Another related approach by Harvill et al.~\cite{harvill2022frame} uses a frame-wise dysfluency detection model for stutter correction, utilising bidirectional LSTMs and fully connected layers.
The method was evaluated for stutter correction solely by non-experts on Amazon Mechanical Turk, using just 20 samples per dysfluency class and lacking rigorous objective accuracy assessment, making it challenging to determine its broader generalisability.

Thus, we introduce ``StutterCut'', a graph-based soft-constrained clustering method~\cite{basu2008constrained}, designed to segment dysfluent regions \textit{without relying on strong dysfluency labels}.
In our approach, embeddings from overlapping windows of an utterance serve as graph nodes, and edges represent the similarity between these embeddings.
We use normalised cut (N-Cut) to partition the graph, whose dual consideration prevents over-fragmentation and ensures balanced segmentation~\cite{wang2023tokencut}.
To further refine node connectivity, we integrate knowledge from a pseudo-oracle classifier trained on weak dysfluency labels.
This classifier mimics expert guidance by imposing soft segmentation constraints, with its influence adjusted by predictive uncertainty via Monte Carlo dropout entropy~\cite{meyer2021uncertainty}.
This ensures that only reliable predictions guide segmentation.

Additionally, we extend the weakly labelled FluencyBank~\cite{ratner2018fluency} corpus by introducing FluencyBank++, which includes strong labels for four dysfluency types: prolongation, repetition, interjection, and block, offering a more authentic benchmark than synthetic datasets.
Experiments on both synthetic and real-world stuttering datasets demonstrate that StutterCut effectively segments dysfluent regions, outperforming existing methods in dysfluency segmentation.
The code and the FluencyBank++ dataset will be publicly released to support reproducibility and further research\footnote{github.com/suhitaghosh10/stuttercut}.

\section{Normalised Cut (N-Cut)}\label{sec:ncut}
Consider an undirected, weighted and fully-connected graph \(G = (\mathcal{N}, \mathcal{E})\), where \(\mathcal{N}\) is the set of \(N\) nodes and \(\mathcal{E}\) is the set of edges. Let \(\mathbf{W}\) be the similarity matrix, where \(\mathbf{W}_{i,j}\) is the weight of the edge between the nodes \(n_i\) and \(n_j\), and let \(\mathbf{D}\) be a diagonal matrix with entries \(\mathbf{D}_i = \sum_j \mathbf{W}_{i,j}\). N-Cut~\cite{shi2000normalized} splits \(\mathcal{N}\) into two disjoint subsets, \(\mathcal{S}_1\) and \(\mathcal{S}_2\). The membership of the nodes is given by \(\mathbf{y} \in \{1,-b\}^N\), where \(b\) is a constant that ensures a balanced partition, preventing one subset from dominating the other~\cite{shi2000normalized}. N-Cut aims to partition the graph to minimise edges between subsets while preserving strong connectivity within each subset.
It can be formulated as a generalised eigenvalue problem, whose \emph{eigenvectors define the partition}~\cite{shi2000normalized}. Specifically, one solves: \(\mathbf{D}^{-\frac{1}{2}}(\mathbf{D}-\mathbf{W})\mathbf{D}^{-\frac{1}{2}} = \lambda \mathbf{z}\),
where \(\mathbf{D}-\mathbf{W}\) is the graph Laplacian, known to be positive semidefinite, and \(\mathbf{z} = \mathbf{D}^{-\frac{1}{2}}\mathbf{y}\), ensuring efficient eigen-decomposition~\cite{shi2000normalized}. The smallest eigenvector, \(\mathbf{z}_0 = \mathbf{D}^{\frac{1}{2}}\mathbf{1}\) (corresponding to \(\lambda = 0\)), is trivial and not useful for partitioning. Instead, the second smallest eigenvector, \(\mathbf{z}_1\), also known as the \textbf{Fiedler vector}, provides the N-Cut solution (after substituting \(\mathbf{z} = \mathbf{D}^{\frac{1}{2}}\mathbf{y}\)).

\vspace{-0.5em}
\begin{equation}
\mathbf{y}_1 = \text{argmin}_{\mathbf{y}^T\mathbf{D}\mathbf{1}=0}\frac{\mathbf{y}^T(\mathbf{D}-\mathbf{W})\mathbf{y}}{\mathbf{y}^{T}\mathbf{Dy}}
\end{equation}
 
\noindent where the two $\mathcal{S}_1$ and $\mathcal{S}_2$ clusters are obtained by taking the sign or applying a threshold to $\mathbf{y}_1$~\cite{shi2000normalized}.
Finer clusters can also be obtained using higher-order eigenvectors~\cite{wang2023speech}.

\section{StutterCut} \label{sec:sc}
We propose StutterCut, a method that segments dysfluent regions using only utterance-level (weak) dysfluency labels.
Specifically, we refine the similarity matrix derived from Fiedler vectors by incorporating guidance from a pseudo-oracle classifier $\mathcal{C}$.
The classifier is trained on weak labels, serving as a proxy for human expertise.
This classifier introduces soft constraints, indicating which segments within an utterance are more likely to be grouped or separated.
Unlike hard-constrained clustering~\cite{basu2008constrained}, our soft constraints are incorporated by modulating the similarity matrix, enabling three key benefits:~(1) independence from strong dysfluency labels, which are costly and time-consuming for experts to annotate,~(2) flexibility to ignore misleading constraints, and~(3) freedom to choose any clustering procedure.
Our method comprises five stages, as shown in Figure~\ref{fig:sc}:
\vspace{0.2em}

\noindent\textbf{1.~Initial Graph Generation:} \label{step1}
We divide an utterance into \(N\) overlapping windows, each of length \(l\) with a stride of \(r\).
For each window, we extract embeddings using a pre-trained foundation model (such as WavLM~\cite{chen2022wavlm}), whose representations capture the subtle acoustic and linguistic nuances essential for dysfluency detection, such as stretched sounds (in prolongations) and silent pauses caused by articulatory difficulty (in blocks).
Each embedding corresponds to a node \(n_i\) in the graph \(G\).
We compute edge weights using the cosine similarity between node embeddings and construct the initial similarity matrix \(\mathbf{W}^{(1)}\) from these weights.
Intuitively, embeddings from the same class (dysfluent or non-dysfluent) tend to be closer in latent space~\cite{ghosh24_interspeech}, resulting in higher similarity values.
\vspace{0.2em}

\noindent\textbf{2.~Knowledge Extraction from Pseudo-Oracle:}
At this stage, we derive another similarity matrix $\mathbf{W}^{(2)}$ from the predictions of the pseudo-oracle classifier $\mathcal{C}$, trained in a multilabel fashion using weak dysfluency labels.
To avoid any data leakage and ensure real-world applicability, $\mathcal{C}$ is trained on a separate dataset (or disjoint partition) from that used in the first stage, where only a pre-trained foundation model is used to extract embeddings.
The similarity matrix $\mathbf{W}^{(2)}$ is derived from the cosine similarity of prediction probabilities, as shown in Equation~\eqref{eqn:P}.
Here, \(\mathbf{P}_0 \in \mathbb{R}^N\) is the probability vector for the non-dysfluent (fluent) class, where \(\mathbf{P}_{i,0}\) denotes the probability of node \(n_i\) being fluent. Similarly, \(\mathbf{P}_{\max} \in \mathbb{R}^N\) is defined by \(\mathbf{P}_{i,\max} = \max_{k=1}^K \mathbf{P}_{i,k}\), capturing the highest predicted probability among the \(K\) dysfluent classes for node \(n_i\).
Since a window may contain both fluent and dysfluent segments, using both probabilities ensures a more informed distinction.
\vspace{-0.5em}
\begin{equation}
\label{eqn:P}
\mathbf{W}^{\text{(2)}} = \mathbf{P}_{0} \mathbf{P}_{0}^T + \mathbf{P}_{\text{max}}  \mathbf{P}_{\text{max}}^T 
\end{equation}

\noindent\textbf{3.~Prior Knowledge Integration from Pseudo-Oracle:}
At this stage, we derive the refined similarity matrix \(\widetilde{\mathbf{W}}\) by fusing \(\mathbf{W}^{\mathrm{(1)}}\) and \(\mathbf{W}^{\mathrm{(2)}}\) according to:
\vspace{-0.5em}
\begin{equation}
\widetilde{\mathbf{W}} =
 \mathbf{W}^{(1)} \odot (\mathbf{I} - \mathbf{D}_\mathrm{M}) + (\mathbf{W}^{(1)} \mathbf{W}^{(2)}) \odot \mathbf{D}_\mathrm{M}
\end{equation}

\noindent where \(\mathbf{I} \in \mathbb{R}^{N \times N}\) is an identity matrix, and \(\mathbf{D}_\mathrm{M} = \operatorname{diag}(\mathbf{M})\) is a diagonal matrix formed from the \(N\)-dimensional mask \(\mathbf{M}\), which controls the contribution of the pseudo-oracle.
The mask value \(\mathbf{M}_i\) represents the confidence of the pseudo-oracle's prediction for node \(n_i\) and is derived as \(\mathbf{M}_i = 1 - \frac{\mathbf{H}_i}{\text{H}_{\max}}\), where \(\mathbf{H}_i\) is the entropy estimated from multiple stochastic forward passes (MC dropout)~\cite{meyer2021uncertainty} and, \(\text{H}_{\max} = \log K\), denotes the maximum possible entropy for \(K\) classes.
The scaling by \(\text{H}_{\max}\) ensures that \(\mathbf{M}_i\) remains within the range \([0,1]\). Finally, to strengthen highly similar connections while maintaining overall graph connectivity, we threshold by setting similarities below \(\tau\) to a small value (\(10^{-5}\)), ensuring that weaker connections are preserved but providing minimal influence.
\vspace{0.2em}

\noindent\textbf{4.~Graph Partition:}
We apply N-Cut to \(\widetilde{\mathbf{W}}\) to produce two clusters: dysfluent and non-dysfluent.
The partitioning is determined using the signs of the Fiedler vector \(\mathbf{y}_1\) or its average value \(\bar{\mathbf{y}}_1\), following the approach in~\cite{wang2023tokencut}.
Formally, \(
\mathcal{S}_1 = \{ n_i \mid \mathbf{y}_{1,i} \le \phi \}, \quad 
\mathcal{S}_2 = \{ n_i \mid \mathbf{y}_{1,i} > \phi \}, \quad \text{where}\  \phi \in \{ 0,\ \bar{\mathbf{y}}_1 \}
\).
Unlike in image-based methods~\cite{wang2023tokencut}, where the cluster with the highest Fiedler vector identifies the salient object due to a clear foreground-background distinction, speech lacks such separation. Instead, we determine the dysfluent cluster as the one with a higher number of dysfluent nodes, identified using \(\mathcal{C}\).

\vspace{0.2em}

\noindent\textbf{5.~Boundary Extraction (BE):}
We derive the boundaries of dysfluencies by merging consecutive dysfluent windows into continuous segments, taking the earliest start time $t_{0}$ and the latest end time $t_{1}$, as shown in Figure~\ref{fig:sc}.
Short gaps ($\leq$~$\eta$~$\si{\milli\second}$) are merged to reduce fragmentation, and isolated short segments ($\leq$~$\eta$~$\si{\milli\second}$) are removed to limit false positives.
\section{FluencyBank++ Dataset}
One of the major challenges in dysfluency segmentation is the scarcity of real-world datasets.
To address this limitation, we extend FluencyBank~\cite{ratner2018fluency} with strong dysfluency labels.
We call our augmented version FluencyBank++.

\noindent\textbf{Annotation Protocol}:
FluencyBank consists of 4,144 three-second clips from 33 interviews (23 male, 10 female speakers), annotated for five dysfluency types: block, interjection, prolongation, sound repetition, and word repetition.
We combine sound, word, syllable, and phrase repetitions into a single category: \textit{repetitions}.
Further, we filter out clips lacking speech or containing music and retain only those annotated as dysfluent by at least two annotators.

Annotating dysfluency boundaries is challenging because transitions between dysfluencies are often fuzzy.
For example, interjections such as ``um" can be repeated and may blend with sound repetitions or prolongations, and multiple dysfluency types frequently occur together~\cite{bayerl23_interspeech}.
To improve annotation quality, we employ a multi‐stage strategy with five experts: two dysfluency specialists for the initial stages and three speech pathologists for final refinement.
The three stages are:

\noindent\textbf{1.~Utterance-Level Correction:} Original three-second clips are extended to five seconds (adding one second on each side) to avoid truncating dysfluencies, especially partial word repetitions, reported in~\cite{changawala2024whister}. Two annotators review these extended clips and revise the weak labels, adjusting for any shifts introduced due to the extension.

\noindent\textbf{2.~Frame-Level Boundary Annotation:} 
    Both annotators independently mark start and end points for each dysfluency in the clips approved in the previous step. For utterances with multiple dysfluencies (e.g., block and interjection), each type is presented separately, and the annotator is explicitly informed which dysfluency type to mark. This approach ensures unbiased and structured boundary annotation.

\noindent\textbf{3.~Boundary Refinement and Consensus:} Three speech pathologists review the dysfluency boundaries from the previous stage, ensuring no regions are missed.
    A voting system was implemented, where the frames marked as dysfluent by at least two annotators were finally labelled as dysfluent.
\vspace{0.2em}

\noindent\textbf{Dataset Statistics}:
FluencyBank++ consists of 3,017 five-second clips, which is 72.8\% of the original data after filtering.
Table~\ref{tab:fluencybank_summary} provides an overview of each dysfluency class, including the number of clips and duration ranges.
To assess annotation reliability, we calculated Fleiss’ Kappa over \SI{0.1}{\second} frames following the procedure in~\cite{wang2023speech}.
The Fleiss’ Kappa score improved from 0.47 (moderate agreement) in the second stage to 0.71 (strong agreement) in the final stage, demonstrating that the multi-level process enhanced annotation reliability.
\vspace{-0.5em}
\begin{table}[h]
\centering
\vspace{-0.5em}
\caption{Overview of the FluencyBank++ dataset.}
\vspace{-0.5em}
\label{tab:fluencybank_summary}
\resizebox{0.47\textwidth}{!}{

\begin{tabular}{lccc}
\toprule
\textbf{Dysfluency} & \textbf{\#Clips} & \textbf{Minimum Duration (s)} & \textbf{Maximum Duration (s)} \\
\midrule
Interjection  & 1,130 & 0.12 & 1.88 \\
Repetition    &   921 & 0.20 & 4.99 \\
Block         &   530 & 0.23 & 4.20 \\
Prolongation  &   436 & 0.41 & 3.95 \\
\bottomrule
\end{tabular}
}
\vspace{-1em}
\end{table}
\section{Experiment and Results}

\textbf{Datasets}:
All speech recordings are resampled to 16 kHz.
For our experiments, we consider FluencyBank++, our newly introduced dataset, along with two English stuttering datasets: VCTK-TTS~\cite{zhou2024yolo} and Sep-28K~\cite{lea2021sep}.
VCTK-TTS is a synthetic, strong-labelled dataset extending VCTK~\cite{yamagishi2019cstr}, comprising utterances from 109 native English speakers with various accents and artificially inserted dysfluencies such as repetitions, blocks, missing words, and prolongations.
Sep-28K is a large-scale dysfluency dataset having the same weak dysfluency labels as in FluencyBank.
Since Sep-28K and FluencyBank++ do not include `missing words', we exclude this dysfluency class from our analysis.
\vspace{0.2em}

\noindent\textbf{Baselines}:
We compare StutterCut against four baselines: YOLO-Stutter~\cite{zhou2024yolo}, the current state-of-the-art in dysfluency segmentation; Harvill et al.~\cite{harvill2022frame}, a stutter-correction model closely related to dysfluency segmentation; and two variants of Whister~\cite{changawala2024whister}, a classifier that detects the type of dysfluency by leveraging Whisper representations as input features.
One variant of Whister, Whister\(_{\text{ML}}\), is trained using focal loss in a multi-label classification setup with sigmoid activation instead of softmax, while the other, Whister\(_{\text{MC}}\), follows the multi-class classification approach with weighted cross-entropy loss as described in~\cite{changawala2024whister}.
Since Whister-based models are trained on weak dysfluency labels, they cannot provide explicit boundary information.
Therefore, we extract boundaries by applying the boundary extraction approach (see Section~\ref{sec:sc}) after predicting dysfluency over overlapping windows.
Notably, Whister\textsubscript{ML} also serves as the pseudo-oracle $\mathcal{C}$ in our method.
\vspace{0.2em}

\noindent\textbf{Data Selection and Split}: The Whister-based models and the method by Harvill et al.~\cite{harvill2022frame} are trained on weak labels.
We train these models on VCTK-TTS and Sep-28K, excluding LibriStutter~\cite{libristutter} since its unnatural dysfluencies have been shown to degrade performance in our initial experiments and prior work~\cite{changawala2024whister}.
We use the same data split for training and validation for all models, with no speaker overlap, ensuring fair evaluation.
For Sep-28K, we follow the splits from~\cite{changawala2024whister}, using 1,400 clips from two podcasts (``I Stutter So What'' and ``My Stuttering Life'') for validation and the remainder for training.
However, Sep-28K is excluded from the dysfluency segmentation evaluation due to the absence of strong labels.
For VCTK-TTS, we use 118,085 clips from 60 speakers for training, 12 speakers for validation and the rest for evaluation.
We use the entire FluencyBank++ dataset for evaluation.
\vspace{0.2em}

\noindent\textbf{Training and Hyperparameters}: The models trained on weak labels use the Adam optimiser (\(\beta_1 = 0.9\), \(\beta_2 = 0.999\)) with a learning rate of \(10^{-4}\). 
They are trained with a batch size of 1024 for 100 epochs, halving the learning rate if the validation loss stagnates for 2 epochs, 
following~\cite{changawala2024whister}. 
Training each model takes \(\approx \SI{48}{\hour}\) on an Nvidia A100 (\(\SI{80}{\giga\byte}\)) GPU. 
For YOLO-Stutter, we extract speech encodings using the pre-trained VITS~\cite{kim2021conditional} model, 
obtain transcriptions with WhisperX~\cite{bain23_interspeech}, and perform segmentation using a model pre-trained on VCTK-TTS, 
following the set-up in~\cite{zhou2024yolo}.

For StutterCut, we set a similarity threshold of \(\tau = 0.25\). 
We employ overlapping windows of length \(l = \SI{0.75}{\second}\) and a stride of \(r = \SI{0.1}{\second}\), 
empirically determined on the validation set. 
This configuration provides high temporal resolution for detecting brief interjections 
while preserving sufficient context for longer dysfluencies such as word repetitions. 
The \SI{0.1}{\second} stride creates overlapping windows, allowing the method to capture subtle transitions more effectively and enhance boundary detection.
MC dropout-based uncertainty measures are computed from 100 stochastic forward passes.
\vspace{0.2em}

\noindent\textbf{Inference and Computational Efficiency}:
Using $l = \SI{0.75}{\second}$ and $r = \SI{0.1}{\second}$, StutterCut generates dysfluency boundaries in about \(\SI{2.2}{\second}\) for a \(\SI{1}{\minute}\) utterance. Our method uses Whister\(_{\text{ML}}\) as the pseudo-oracle, which has 7.1 million trainable parameters, less than YOLO-Stutter's 33 million~\cite{zhou2024yolo}.
\vspace{0.2em}

\noindent\textbf{Evaluation Metrics:}
We evaluate performance using three metrics: time-F1 (t-F1), time-recall (t-recall), and onset error.
T-F1 and t-recall~\cite{zhou2024yolo} assess boundary alignment by computing the Intersection over Union (IoU) between predicted and ground-truth intervals, considering a detection true positive if \(\text{IoU} > 0.5\).
T-F1 balances pre-
\begin{table}[!h]

    \centering
    \caption{Performance of methods using different embeddings and clustering algorithms, with overall T-F1 and t-recall scores reported with 95\% confidence intervals.}
    \vspace{-0.5em}
    \label{tab:ablation}
    \resizebox{0.5\textwidth}{!}{

    \begin{tabular}{l l c c | c c}
    \toprule
         {\textbf{\specialcell{0.25cm}{Embedding}}}& 
         {\textbf{\specialcell{0.25cm}{Method}}}& 
         \multicolumn{2}{c|}{\textbf{FluencyBank++}}&
         \multicolumn{2}{c}{\textbf{VCTK-TTS}}\\
         \cline{3-6}
         & &\raisebox{-0.4ex}{\textbf{T-F1 $\uparrow$}} &\raisebox{-0.4ex}{\textbf{T-recall $\uparrow$}} 
         &\raisebox{-0.4ex}{\textbf{T-F1 $\uparrow$}} 
         &\raisebox{-0.4ex}{\textbf{T-recall $\uparrow$}} \\
         \midrule

\multirow{4}{*}{\textbf{Whisper}} &K-Means &59.9 $\pm$ 4.5 &80.9 $\pm$ 4.6  &78.5 $\pm$ 3.8 &80.8 $\pm$ 4.3\\

&FuzzyCMeans &60.7 $\pm$ 4.6 &81.2 $\pm$ 4.6 &78.5 $\pm$ 3.9 &80.6 $\pm$ 4.4 \\

&Fiedler\,[\( \geq \mathbf{\bar{y}_{1}} \)] &63.0 $\pm$ 4.1 &82.9 $\pm$ 4.2 &79.4 $\pm$ 3.4 &\textbf{100.0} $\pm$ 0.0\\

&\textbf{Fiedler}\,[\( \mathbf{y}_{1} \geq 0 \)] &\textbf{63.5} $\pm$ 3.9 &\textbf{83.2} $\pm$ 4.6 &\textbf{82.7} $\pm$ 3.5 &88.0 $\pm$ 3.7\\
\midrule

\multirow{4}{*}{WavLM} &K-Means &59.7 $\pm$ 3.9 &80.7 $\pm$ 4.1 &75.1 $\pm$ 3.0 &86.9 $\pm$ 4.0\\

 &FuzzyCMeans &61.9 $\pm$ 4.5 &81.3 $\pm$ 4.6 &75.4 $\pm$ 2.7 &87.0 $\pm$ 4.0\\
 
 &Fiedler\,[\( \geq \mathbf{\bar{y}_{1}} \)] &63.1 $\pm$ 4.8 &82.6 $\pm$ 4.7 &76.4 $\pm$ 1.0 &\textbf{100.0} $\pm$ 0.0 \\
 
 &Fiedler\,[\( \mathbf{y}_{1} \geq 0 \)] &63.2 $\pm$ 4.7 &82.8 $\pm$ 5.0 &76.8 $\pm$ 3.4 &88.4 $\pm$ 4.6\\
 
\midrule





\multirow{4}{*}{Wav2Vec2} &K-Means &56.3 $\pm$ 4.5 &79.4 $\pm$ 4.7 &71.8 $\pm$ 3.9 &82.0 $\pm$ 4.0\\

 &FuzzyCMeans &59.4 $\pm$ 4.4 &80.6 $\pm$ 4.3 &72.0 $\pm$ 3.9 &81.7 $\pm$ 4.1\\
 
 &Fiedler\,[\( \geq \mathbf{\bar{y}_{1}} \)] &59.5 $\pm$ 4.5 &82.2 $\pm$ 4.7 &72.3 $\pm$ 3.5 &\textbf{100.0} $\pm$ 0.0\\
 
 &Fiedler\,[\( \mathbf{y}_{1} \geq 0 \)] &60.2 $\pm$ 4.7 &82.0 $\pm$ 4.9 &72.9 $\pm$ 3.8 &86.0 $\pm$ 3.9\\

\bottomrule
         
    \end{tabular}%
}
    \label{tab:methods}
    \vspace{-0.6em}
\end{table}

\begin{table*}[!ht]
    \centering
    \caption{95\% confidence intervals are reported for all models. The first four methods are baselines, and the remaining entries are Stutter-Cut variants: SCut (using classifier guidance weighted by MC dropout entropy), SCut\,[\(\mathbf{\textit{M}}_{\mathbf{\textit{P}} \geq 0.5}\)] (using classifier guidance weighted by MC dropout probability), SCut\,[–\(\mathbf{\textit{M}}\)] (using classifier guidance without mask), and SCut\,[–\(\mathbf{\textit{M}} - \mathcal{C}\)] (pure N-Cut). ``Prolong.'' (prolongation), ``Rep.'' (repetition), ``Interj.'' (interjection), ``Block'', ``S.Rep.'' (sound repetition), and ``W.Rep.'' (word repetition) denote dysfluency-class-wise t-F1 scores. T-F1, t-recall, and onset error (Onset E., measured in seconds) represent the overall scores.}
    \vspace{-0.3em}
    \label{tab:final_results}
    \resizebox{\textwidth}{!}{

    \begin{tabular}{@{}l c c c c c c c c c c c c c c@{}}
    \toprule
         
         & \multicolumn{7}{c}{\textbf{\large FluencyBank++}} 
& \multicolumn{7}{c}{\textbf{ \large VCTK-TTS}} \\
\cmidrule(lr){2-8}\cmidrule(lr){9-15}
\textbf{\large Method} 
& \textbf{\large Prolong. $\uparrow$} & \textbf{\large Rep. $\uparrow$} & \textbf{\large Interj. $\uparrow$} & \textbf{\large Block $\uparrow$} & \textbf{\large T-F1 $\uparrow$} & \textbf{\large T-recall $\uparrow$} & \textbf{\large Onset E. $\downarrow$}
& \textbf{\large Prolong. $\uparrow$} & \textbf{\large S. Rep $\uparrow$} & \textbf{\large W. Rep $\uparrow$} & \textbf{\large Block $\uparrow$} & \textbf{\large T-F1 $\uparrow$} & \textbf{\large T-recall $\uparrow$} & \textbf{\large Onset E. $\downarrow$} \\
\midrule
        
        YOLO-Stutter~\cite{zhou2024yolo} &60.2 $\pm$ 4.5 &59.5 $\pm$ 3.3 &44.2 $\pm$ 5.1 &\textbf{63.4} $\pm$ 8.6 &56.8 $\pm$ 3.4 &87.1 $\pm$ 4.7 &0.5 $\pm$ 0.6 &53.6 $\pm$ 3.4 &59.1 $\pm$ 5.4 &\textbf{65.6} $\pm$ 7.4 &56.8 $\pm$ 4.5 &58.8 $\pm$ 3.7 &\textbf{100.0} $\pm$ 0.0 &1.4 $\pm$ 0.7\\

         Harvill et al.~\cite{harvill2022frame} &10.2 $\pm$ 10.9 &39.5 $\pm$ 16.7 &23.2 $\pm$ 14.9 &13.4 $\pm$ 11.6 &21.6 $\pm$ 14.8 &47.1 $\pm$ 13.7 &1.5 $\pm$ 0.6 &21.4 $\pm$ 23.7 &29.2 $\pm$ 12.3 &25.8 $\pm$ 10.8 &24.9 $\pm$ 13.7 &25.3 $\pm$ 13.4 &67.0 $\pm$ 12.3 &1.7 $\pm$ 0.7\\

        $\mathrm{Whister}_{\text{MC}}$~\cite{changawala2024whister} &43.5 $\pm$ 6.3 &35.4 $\pm$ 7.3 &54.1 $\pm$ 6.5 &41.9 $\pm$ 4.6 &43.7 $\pm$ 5.7 &55.5 $\pm$ 6.7 &1.4 $\pm$ 0.5 &27.4 $\pm$ 8.7 &37.2 $\pm$ 11.3 &34.8 $\pm$ 10.8 &47.9 $\pm$ 13.4 &36.8 $\pm$ 7.4 &77.0 $\pm$ 9.1 &1.3 $\pm$ 1.1\\

        $\mathrm{Whister}_{\text{ML}}$ &\textbf{67.5} $\pm$ 4.3 &\textbf{65.8} $\pm$ 4.4 &\textbf{57.3} $\pm$ 4.5 &53.8 $\pm$ 4.7 &\textbf{61.1} $\pm$ 4.8 &\textbf{88.5} $\pm$ 4.3 &\textbf{0.4} $\pm$ 0.5 &\textbf{63.1} $\pm$ 4.2 &\textbf{77.6} $\pm$ 4.1 &65.0 $\pm$ 4.9 &\textbf{76.5} $\pm$ 4.2 &\textbf{70.6} $\pm$ 4.7 &86.7 $\pm$ 4.1 &\textbf{0.4} $\pm$ 0.3 \\
        
        \midrule
        \textbf{SCut} &\textbf{73.2} $\pm$ 4.0 &\textbf{68.8} $\pm$ 4.3 &\textbf{64.5 }$\pm$ 4.5 &\textbf{70.6} $\pm$ 4.4 &\textbf{69.3} $\pm$ 4.9 &\textbf{89.2 }$\pm$ 4.7 &\textbf{0.3} $\pm$ 0.5 &\textbf{84.7} $\pm$ 3.5 &\textbf{91.5 }$\pm$ 2.2 &\textbf{87.6} $\pm$ 3.1 &\textbf{87.6}$\pm$ 3.3 &\textbf{87.2} $\pm$ 3.1 &\textbf{92.6} $\pm$ 3.2 &\textbf{0.2} $\pm$ 0.4\\ 

          SCut\,[$\mathbf{M}_{\text{P}\geq0.5}$] &68.7 $\pm$ 4.3 &65.5 $\pm$ 4.4 &57.2 $\pm$ 5.7 &63.3 $\pm$ 4.3 &63.7 $\pm$ 4.4 &86.6 $\pm$ 4.5 &0.4 $\pm$ 0.5 &84.4 $\pm$ 3.0 &90.2 $\pm$ 2.1 &84.7 $\pm$ 3.7 &87.3 $\pm$ 3.0 &86.6 $\pm$ 3.3 &91.6 $\pm$ 3.3 &\textbf{0.2} $\pm$ 0.4\\
          
          SCut\,[\textminus $\mathbf{M}$] &68.6 $\pm$ 4.2 &63.0 $\pm$ 4.2 &57.4 $\pm$ 4.4 &64.7 $\pm$ 4.0 &63.4 $\pm$ 1.1 &83.2 $\pm$ 2.2 &0.3 $\pm$ 0.4 &83.5 $\pm$ 3.5 &83.0 $\pm$ 3.3 &81.2 $\pm$ 3.1 &83.0 $\pm$ 3.7 &82.7 $\pm$ 3.5 &88.0 $\pm$ 3.7 &0.5 $\pm$ 0.8\\

        SCut\,[\textminus $\mathbf{M}$ \textminus $\mathcal{C}$] &68.2 $\pm$ 4.3 &64.8 $\pm$ 4.5 &54.5 $\pm$ 4.7 &63.6 $\pm$ 4.4 &62.8 $\pm$ 4.9 &85.2 $\pm$ 4.8 &0.4 $\pm$ 0.5 &83.0 $\pm$ 5.5 &89.5 $\pm$ 2.5 &87.3 $\pm$ 5.3 &85.7 $\pm$ 3.2 &86.1 $\pm$ 3.1 &90.1 $\pm$ 3.2 &0.3 $\pm$ 0.4\\

         \bottomrule
    \end{tabular}%
}
    \label{tab:subjective}
    \vspace{-1.1em}
\end{table*}
\noindent cision while t-recall measures the proportion of correctly identified dysfluency segments.
Onset error, defined as the absolute difference between predicted and actual onset times, quantifies the accuracy of dysfluency onset detection.

Dysfluency-class-wise scores are computed by first calculating the metric for each speaker in the test set and then macro-averaging these scores across speakers, which prevents bias from speakers with more data and ensures equal representation~\cite{changawala2024whister}.
Overall performance is obtained by macro-averaging these dysfluency-class-wise scores to mitigate class imbalance, particularly in FluencyBank++, where interjections occur over twice as often as prolongations.
This approach treats all dysfluency classes equally and aligns with prior work on unbalanced datasets~\cite{li20v_interspeech,cornell23_chime}.
\vspace{0.2em}

\noindent\textbf{Selection of Embeddings and Clustering Method}:
We explore different combinations of clustering methods and speech embeddings to determine the optimal setup for StutterCut.
Specifically, we compare the graph-based N-Cut method with traditional clustering approaches (K-means and Fuzzy C-means), leveraging embeddings from foundation models known for capturing dysfluency representations~\cite{changawala2024whister,wagner2024large,shih2024self}.
We consider the embeddings from the ASR-based foundation model Whisper alongside the self-supervised models WavLM Large~\cite{chen2022wavlm} and Wav2Vec2~\cite{baevski2020wav2vec}.
Following~\cite{changawala2024whister}, we extract Whisper embeddings from all encoder layers, while for Wav2Vec2 and WavLM we use the 12\textsuperscript{th} and 20\textsuperscript{th} layers, respectively, as these better capture the nuances essential for dysfluency detection~\cite{wagner2024large,shih2024self}.
Finally, we evaluate two Fiedler threshold variants: one using a sign threshold (Fiedler\,[\(\mathbf{y}_1 \geq 0\)]) and the other using the mean Fiedler vector as the threshold (Fiedler\,[\(\geq \bar{\mathbf{y}}_1\)]).

Table~\ref{tab:ablation} shows that Fiedler-based methods consistently outperform K-Means and Fuzzy C-Means in t-F1 and t-recall.
Regardless of the type of embeddings, Fiedler\,[\( \geq \mathbf{\bar{y}_{1}} \)] achieves the highest mean t-recall (100.0\%) on VCTK-TTS but with lower precision due to synthetic noise misclassification.
In contrast, Fiedler\,[\( \mathbf{y}_{1} \geq 0 \)] reduces over-detection, improving precision at a slight cost to mean t-recall.
As the variations in onset error were minimal (\SIrange{0.3}{0.8}{\second}) for both datasets, detailed results are omitted from the table.
Ultimately, we consider Whisper embeddings with Fiedler\,[\( \mathbf{y}_{1} \geq 0 \)], as they provide the best mean t-F1 for both of the datasets.
\vspace{0.2em}

\noindent\textbf{Comparison to Baselines}:
Table~\ref{tab:final_results} demonstrates the effectiveness of StutterCut. Our best variant, SCut, using classifier guidance weighted by MC dropout entropy, achieves the highest overall performance on both FluencyBank++ and VCTK-TTS. Among baselines, YOLO-Stutter shows high mean t-recall (87.1\% in FluencyBank++, 100.0\% in VCTK-TTS) but low overall mean t-F1 (56.8\% in FluencyBank++, 58.8\% in VCTK-TTS) due to over-detection of dysfluencies. The method by Harvill et al.~\cite{harvill2022frame} struggles significantly, especially in prolongation detection, making it the weakest.
Whister\textsubscript{ML}, also used as the pseudo-oracle in our approach, outperforms other baselines but has difficulty with the block (mean t-F1: 53.8\%) and interjection detection (mean t-F1: 57.3\%) on FluencyBank++.
SCut improves overall mean t-F1 over Whisper\textsubscript{ML} by 13.4\% on FluencyBank++ and 24.1\% on VCTK-TTS.
As both methods use Whisper embeddings, the improvement stems from the synergy of graph-based clustering and classifier-guided constraints.
Nonetheless, StutterCut struggles with interjections (e.g., ``um'', ``like'', ``aah'') that lack pauses or articulatory struggle, often failing to detect them.
Additionally, blocks are often misclassified as pauses, an issue noted in prior dysfluency classification works~\cite{changawala2024whister}.

\noindent\textbf{Ablation Study}:
We explore an alternative uncertainty-weighting variant, SCut\,[\(\mathbf{M}_{p \geq 0.5}\)], where confidence is based on the mean probability computed over stochastic MC passes~\cite{meyer2021uncertainty}. However, it proves less effective than entropy-based weighting, as softmax probabilities tend to be overconfident~\cite{guo2017calibration}.

Retaining classifier guidance but removing the entropy uncertainty mask (SCut\,[\(-\mathbf{M}\)]) lowers the overall mean t-F1 of our best-performing variant SCut from 69.3\% to 63.4\% on FluencyBank++ and from 87.2\% to 82.7\% on VCTK-TTS.
Disabling both classifier constraints and uncertainty weighting (SCut\,[\(-\mathbf{M}-\mathcal{C}\)]), which corresponds to the pure N-Cut method, further degrades performance, reducing mean t-F1 by 10\% and t-recall by 5\% on FluencyBank++, reaffirming the necessity of uncertainty-weighted classifier guidance.
Interestingly, in VCTK-TTS, the pure N-Cut method (SCut\,[\(-\mathbf{M}-\mathcal{C}\)]) already surpasses Whister\textsubscript{ML}, with minimal gains from MC entropy-weighted \(\mathcal{C}\) priors.
This is likely due to the synthetic nature of VCTK-TTS dysfluencies, where predictable patterns and artificial cues simplify segmentation, leading to consistently higher t-F1 and t-recall compared to the FluencyBank++ dataset.

\section{Conclusion and Future Work}
We introduce StutterCut, a semi-supervised graph-based framework for dysfluency segmentation that operates without strong labels or transcriptions, making it more applicable to real-world scenarios.
Our method strikes an optimal balance between recall and precision, outperforming conventional clustering and supervised deep learning baselines on real and synthetic datasets.
We also present FluencyBank++, a dataset with frame-level labels for four dysfluency types that, although limited in size for training, provides a realistic benchmark compared to synthetic datasets.

Future work will explore alternatives to weakly supervised classifiers for cluster identification, such as higher-order statistics from the Fiedler vector.
We also aim to develop adaptive window sizing, as our preliminary observations indicate that the same temporal resolution is not optimal for all dysfluency types (e.g., longer windows for word repetitions and shorter ones for interjections).
Additionally, we plan to incorporate acoustic features to better distinguish interjections and blocks from fluent speech and to accurately separate overlapping dysfluencies.
Finally, we aim to evaluate StutterCut on multilingual stuttering corpora to further establish its generalisability.

\section{Acknowledgements}
This research has been supported by the Federal Ministry of
Education and Research of Germany through the projects Medinym
(focused on AI-based anonymisation of personal patient data in
clinical text and voice datasets) and Anti-Stotter (overcoming
stuttering with AI).

\bibliographystyle{IEEEtran}
\bibliography{mybib}

\end{document}